\documentclass[letterpaper]{article}
\usepackage{imav}
\usepackage{times}
\usepackage{graphicx}
\usepackage{float} 

\usepackage{wasysym}

\title{Bumblebees Exhibit Adaptive Flapping Responses to Air Disturbances}

\author{T. Jakobi*\thanks{Email address(es): t.jakobi@unsw.edu.au}, S. Watkins**, A. Fisher**, and S. Ravi* \\ *University of New South Wales and **RMIT University}

\begin{document}

\maketitle
\thispagestyle{empty} 


\begin{abstract}
Insects excel in trajectory and attitude handling during flight, yet the specific kinematic behaviours they use for maintaining stability in air disturbances are not fully understood. This study investigates the adaptive strategies of bumblebees when exposed to gust disturbances directed from three different angles within a plane cross-sectional to their flight path. By analyzing characteristic wing motions during gust traversal, we aim to uncover the mechanisms that enable bumblebees to maintain control in unsteady environments. We utilised high-speed cameras to capture detailed flight paths, allowing us to extract dynamic information.  Our results reveal that bees make differential bilateral kinematic adjustments based on gust direction: sideward gusts elicit posterior shifts in the wing closest to the gust, while upward gusts trigger coordinated posterior shifts in both wings. Downward gusts prompted broader flapping and increased flapping frequencies, along with variations in flap timing and sweep angle. Stroke sweep angle was a primary factor influencing recovery responses, coupled with motion around the flap axis. The adaptive behaviours strategically position the wings to optimize gust reception and enhance wing-generated forces. These strategies can be distilled into specific behavioural patterns that can be analytically modelled to inform the design of robotic flyers. We observed a characteristic posterior shift of wings when particular counteractive manoeuvres were required. This adjustment reduced the portion of the stroke during which the wing receiving gust forces was positioned in front of the centre of gravity, potentially enhancing manoeuvrability and enabling more effective recovery manoeuvres. These findings deepen our understanding of insect flight dynamics and offer promising strategies for enhancing the stability and manoeuvrability of MAVs in turbulent environments.
\end{abstract}

\section{Introduction} \label{section:introduction}

Insects have mastered the art of nimble control over their trajectory and attitude during flight, navigating both natural and urban environments with remarkable precision. These flight characteristics have inspired extensive studies into flapping wings, advancing our understanding of essential flight concepts such as unsteady lift mechanisms \cite{ell1996,dickinson1999wing,sane2003aerodynamics}, control capabilities \cite{SaneDickinson2001, deng2006flapping}, stability \cite{cheng2011translational, Ristroph2013}, and underlying wing functions \cite{wootton1992functional,usherwood2002aerodynamics,zhao2010aerodynamic}. Research on these aspects has provided intelligent inspiration for robotic design \cite{nakata2011aerodynamics}, significantly contributing to the development of Micro Air Vehicles (MAVs).

For smaller-scale flying systems, complex airflow that is highly variable in strength and structure exists ubiquitously and impedes their control performance \cite{watkins06}. Even away from local effects such as wakes of structures and vegetation, wind closer to the ground is highly turbulent due to surface roughness in the atmospheric boundary layer (ABL) \cite{abdulsim}. Animals, particularly insects, use flapping wings to produce the aerodynamic forces necessary for flight. Unsteady wing-air mechanisms such as dynamic stall and wake capture enable flapping wings to operate at low velocities, permitting precise control manoeuvres while hovering. Flapping wings at the low Reynolds numbers relevant to this scale also show positive effects in turbulence, aided by leading-edge vortex (LEV) formation \cite{fisher2013}, and has been shown to help overcome the effects of vortices \cite{Ortega-Jimenez4567,ravibiewe} and gusts \cite{fisheretal,viswanath2010effect}.

Within the demands of insect and MAV flight, the most aerodynamically challenging events are likely the step changes between air scenarios, particularly object approaches, which involve traversing shear layers and structured flows near surfaces. The scale of these local air phenomena are often close to a few characteristic insect-wing dimensions, making these transitions especially demanding. For a flying insect or MAV navigating through one of these wakes, the adverse flow interacting with the wings can be adequately described by a discrete gust containing local flow predominantly in a singular direction. From this, it can be understood that isolated gusts are an unavoidable element of the flight conditions at smaller scales.

Understanding the interplay between normal flapping forward flight and disturbance responses could provide valuable insights into the mechanical effects of certain s during flight, simplifying the situation by reducing it to more basic models. Recent studies demonstrate that organized voluntary body orientation manoeuvres accompany the translatory motions observed among many flying insects during turning, take-off, and landing. They also indicate  the average force periods incident upon the wings, reflecting the higher-order intentions of the bee through the wing outputs, which, when observed instantaneously, can appear as disorganised, chaotic data. During disturbances, the interaction of forces and wing motion become orders more complex, so understanding the average force periods incident upon the wings over longer duration (rather than instantaneous snapshots) could yield more helpful during a disturbance. Simplified models for the role of wing motion during disturbance negotiation could be found by investigating the top-down  and motion effects displayed in high-speed experimental results, which is the focus of this study. 

We extracted the kinematics of bumblebee wings during free flight traversal through strong gusts from three orthogonal directions. By analyzing the rotation angles of the wings, we aimed to understand how wing motion adapts to the direction of the gust incident upon the bee. This analysis provides valuable insights into how flapping wing motions in disturbances can effectively counteract adverse conditions, thereby maintaining stable flight in harsh outdoor airflows. Such understanding enhances our knowledge of insect flight mechanics and could inform the design of more resilient bio-inspired flying robots.

\section{Methodology} \label{section:methodology}

\subsection{Biological Specimen}

Bumblebees are exemplary subjects for studying flight dynamics due to their advanced flapping control capabilities. They perform rapid manoeuvres, utilize sophisticated sensory and control strategies, carry substantial payloads, and function effectively in various weather conditions. Bumblebees are known for their adaptability and resilience, and are proficient learners and navigators, often traveling long distances for food. Their ability to adapt to new environments and execute consistent foraging missions makes them ideal for laboratory observation.

\subsection{Laboratory Environment}

\begin{figure}[hbt]
\centering
\includegraphics[width=1.05\columnwidth]{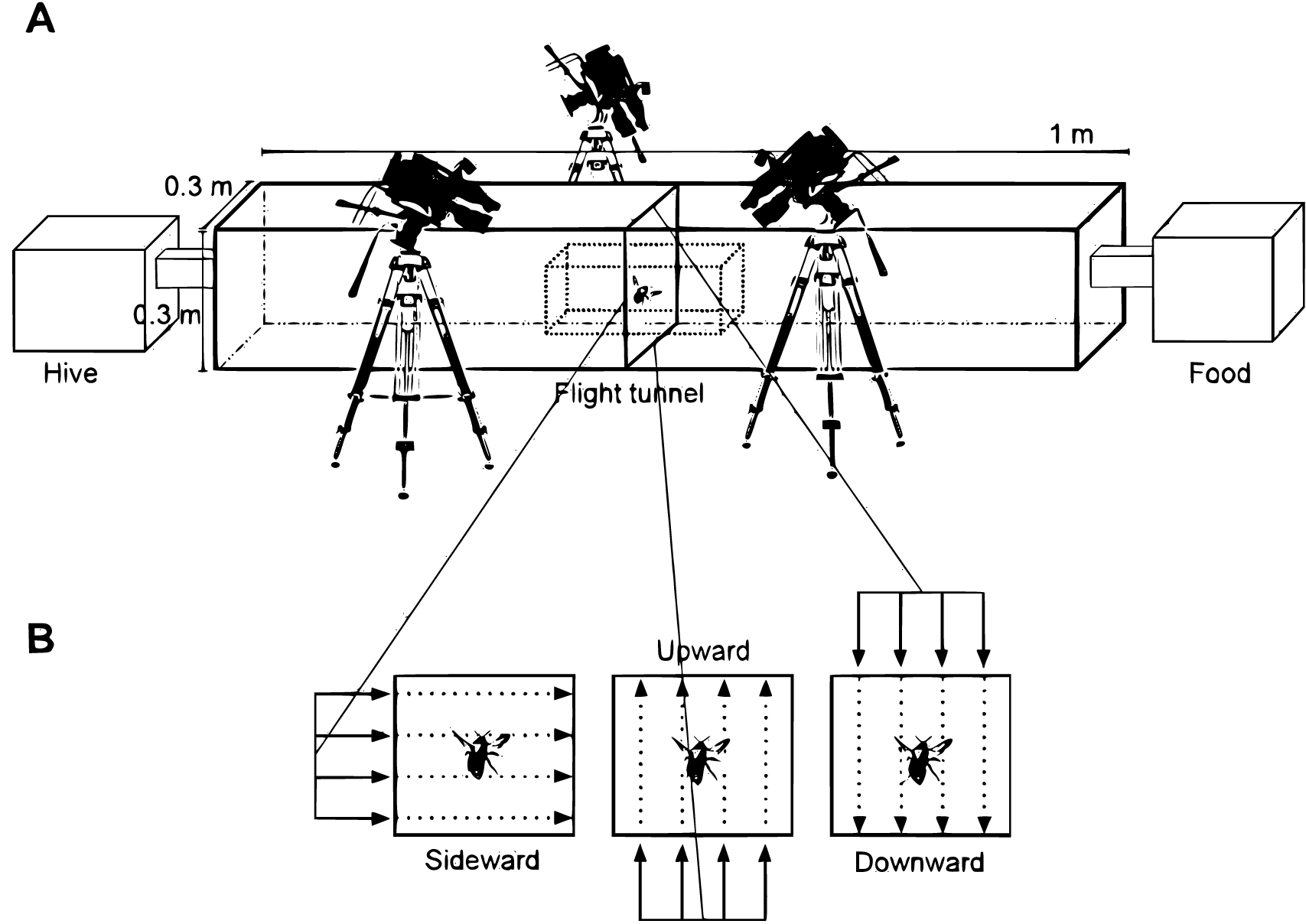}
\caption{Experimental setup for bumblebee flight in gust disturbances. The laboratory environment for bumblebees contained a hive area (left), an interrogation flight tunnel (middle) and a feeding chamber (right). Gusts were made to flow across the cross-section of the interrogation region (B).}
\label{figure:labenv}
\end{figure}

The bumblebees species used in these experiments (\textit{Bombus ignitus}) were maintained under controlled laboratory conditions as shown in Fig. \ref{figure:labenv}A. The bees were sourced from a commercial breeder (Mini Polblack, Koppert, Arysta LifeScience Asia, Japan) as a single hive containing approximately 100 bees. From this hive, around 50 healthy foraging bees were identified and selected for the experiments. 
Data collected from bees in the same experimental setup as described here are published in \cite{jakobi2018bees}. More details on the experimental setup can be obtained by referring to that article.

The selected bees were anesthetized for the attachment of trackable markers. They were housed in a plastic cube measuring approximately 30 cm in each dimension. During a four-day habituation period, the bees were gradually acclimated to the experimental environment. Foraging bees were lured with a sucrose solution with a 1:4 sugar-to-water ratio (by weight) into the flight tunnel where bees could be interrogated (Fig. \ref{figure:labenv}). They were trained to make repeated, consistent trips through the tunnel between the hive and the food source in this setup. The food sources were replenished at the end of each day to ensure continuous traffic to and from the hive. Trained bees flew directly from one end of the tunnel to the other, transporting sugar and pollen loads from the food chamber. Although new bees occasionally made flights through the tunnel, it was evident when a bee was familiar with the environment. Only these learned bees were selected for experimentation.

Bees were anesthetized by placing them individually into a refrigerator for up to 45 minutes. Once anesthetized, small paper markers were affixed to the dorsal surface of the thorax using a combination of wax adhesive and cyanoacrylate glue. After the adhesive dried, the bees were returned to the hive, where they resumed their routine behaviour within minutes. The bees appeared unaffected by the anesthetization process and the thorax-affixed trackable markers.

\subsection{Implementation of Gust Disturbances}

\begin{figure}[hbt]
\centering
\includegraphics[width=0.99\columnwidth]{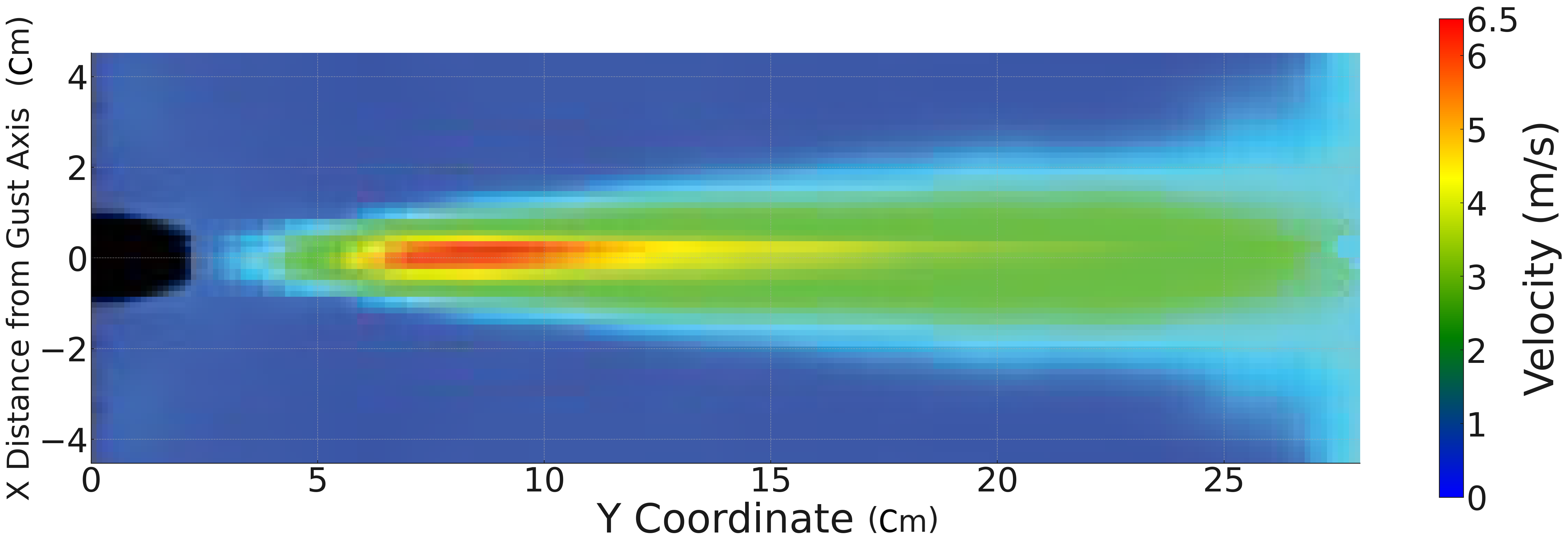}
\caption{PIV image of the gust flow in the tunnel, showing velocities in the middle cross-section of the gust sheet mapped by colour. The example displays a gust oriented sideways with the camera frame looking down on the slice represented in the XY plane. The gust traverses a vertical cross-section (YZ) of the tunnel, moving from left to right over a period of 1 second (an average of 300 frames).}
\label{figure:gustcharac}
\end{figure}

To study the bees' responses, we designed a flight tunnel with a controlled gust disturbance at the midpoint. The disturbance was generated using an air pump connected to an air knife nozzle, which produced a high-speed, 2D air jet across the tunnel's cross-section which challenged the bees without causing them to lose control. The gust velocity was set at 5 $\mathrm{m \, s^{-1}}$, determined by averaging the jet region of the flow between the two shear layers visible in Fig. \ref{figure:coords}. A visual representation of the gust disturbance within the tunnel, captured using Particle Image Velocimetry (PIV) is provided in Fig. \ref{figure:gustcharac}. This technique allowed us to map the gust velocity and flow pattern over time, ensuring that the disturbance was consistent across experiments.

\subsection{3D Localisation and Coordinate Systems}

\begin{figure}[hbt]
\centering
\includegraphics[width=1.2\columnwidth]{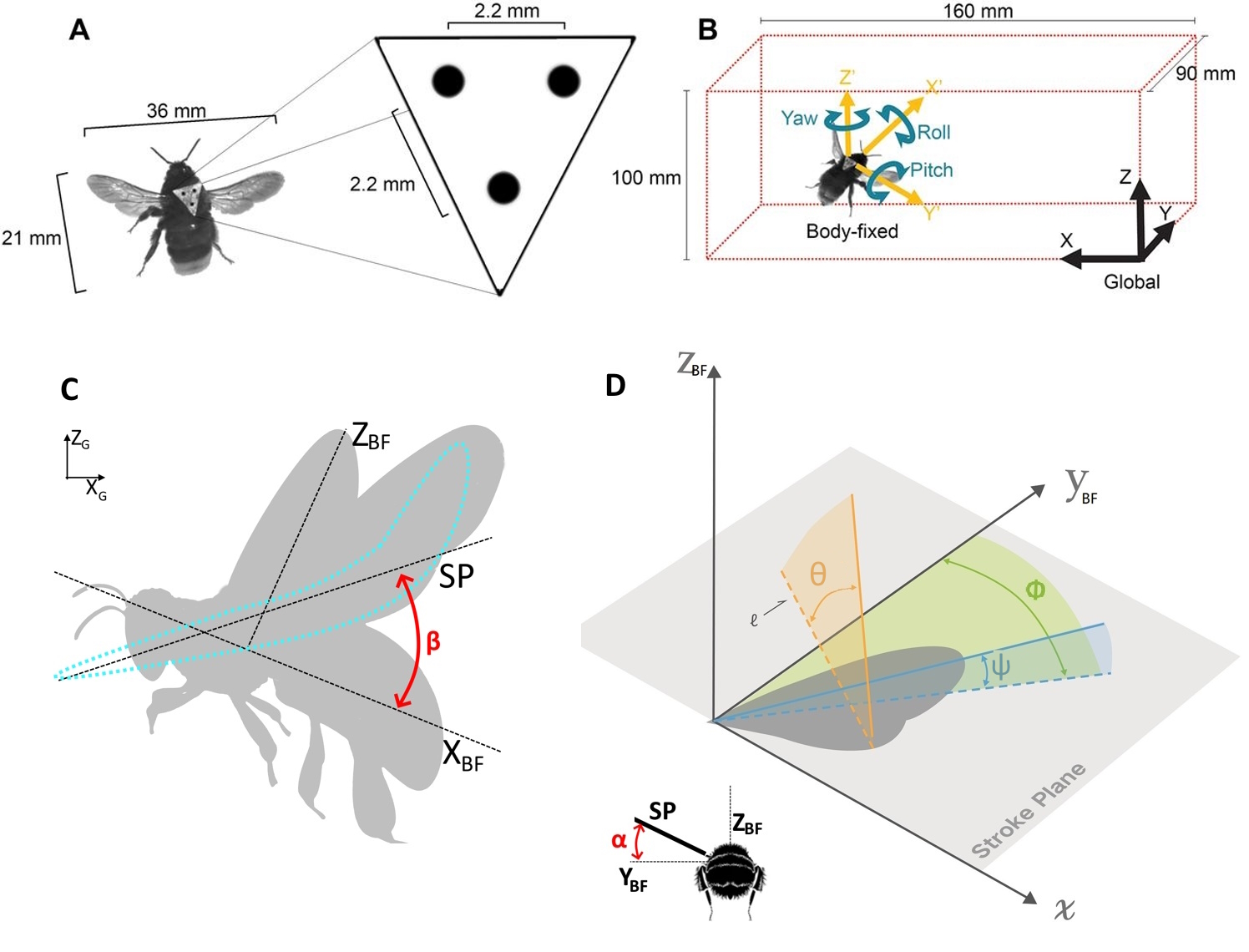}
\caption{(A) The configuration of the trackable markers attached to the thoraxes of bees. (B) The global and body-fixed coordinate system axes used for defining bee motion. (C,D) The layout of the stroke plane relative to the global and body-fixed reference frames. (C) Shows how the wing moves relative to the bee body, with the blue dotted outline showing the path taken by the wing tip of a typical flapping stroke. The stroke plane angles relative to the body-fixed axes are illustrated for reference and were nearly horizontal. (D) Shows the wing angles for sweeping ($\phi$), flapping ($\psi$), and feathering ($\theta$) used to define the wing motion relative to the body-fixed coordinate system.}
\label{figure:coords}
\end{figure}

\begin{table}[h!]
\centering
\caption{Coordinate System Labels and Positive Directions.}
\begin{tabular}{ll}
\hline
\textbf{Label} & \textbf{Positive Direction} \\
\hline
$X_G$ & Forward direction of the tunnel (facing hive side) \\
$Y_G$ & Left-side of the tunnel (facing hive side) \\
$Z_G$ & Top of the tunnel \\
$X_{BF}$ & Anterior direction \\
$Y_{BF}$ & Left-wing lateral direction \\
$Z_{BF}$ & Dorsal direction \\
$Roll$ & Right wing down \\
$Pitch$ & Nose up \\
$Yaw$ & Clockwise looking down \\
\hline
\end{tabular}
\label{tab:coordinates}
\end{table}

We used three Photron high-speed cameras (recording at 2000 fps with a shutter speed of 1/5000) positioned strategically around the tunnel to capture the bees' wing and body motions. To track body kinematics during flight, markers with three black dots in a triangular formation were affixed to the dorsal surface of the bumblebees' thoraxes (Fig. \ref{figure:coords}A). The markers were laser-printed on 80 gsm paper, and the black dots on a white background facilitated subsequent digitization using automated image-feature tracking software \cite{hedrick2008software}. For wing motion tracking, four distinct anatomical landmarks on each wing were identified and tracked across at least two camera views per frame using automated and manual image feature extraction and part recognition algorithms. A 4th order Butterworth filter, chosen for its sharp roll-off characteristics, was applied post-digitization with separate cutoff frequencies tailored to wing and body data. This approach effectively removed noise while preserving the integrity of the recorded motion. The accuracy of automated tracking was validated by manual verification of a subset of frames, ensuring reliable measurement of wing kinematics.

Three coordinate systems were defined. Firstly, the global coordinate system ($XYZ_G$) represents tracked bee coordinates within the flight tunnel (Fig. \ref{figure:coords}B). Second, the body-fixed coordinate system ($XYZ_{BF}$) identifies points relative to the body of the bee (Fig. \ref{figure:coords}B). This coordinate system is aligned with the southern point on the triangular marker, with the $X_{BF}$ axis running longitudinally up the centre of the marker (and the bee body). The $XYZ_{BF}$ is used for calculating body orientation changes and the stroke plane of each wing. Third, the stroke plane system is used for defining the angles of wing motion (Fig. \ref{figure:coords}C and D). For simplicity, the stroke plane coordinate system is considered the same as the body-fixed coordinate system but with offset angles between the marker (body-fixed coordinate system) and the stroke plane applied.

For wing motion, the flap, sweep and feather angles ($\psi$, $\phi$, and $\theta$) were defined in reference to the stroke plane (the mean plane through which the wing tip travels). Only wing sweep and flapping angle were analyzed in detail due to the high sensitivity and complexities involved with minute changes in the feathering angle. The steady flight stroke plane of bees was defined through careful analysis of the wing motion during the phase of flight where no disturbance was present. The offset between the plane of the markers and the bee stroke planes were identified to ensure that the wing kinematics were represented in the relevant frame of reference.

The stroke cycle typically begins and ends at the zero stroke sweep angle, which is parallel to the $Y_{BF}$ axis (with offset applied to the stroke plane) (Fig. \ref{figure:coords}D). However, due to the large variations in translational and rotational wing motions induced by the gusts in this study, the wing sweep angle 
varied significantly during the disturbance phases and independently between each wing. Consequently, it could not reliably define the start and end of the strokes. To uniformly identify the stroke phase during flight for both wings, the peak flapping angle ($\psi_{Max}$), which was more consistent across conditions and still coupled to wing sweep, was used as the reference point for the start of the stroke, with the timing of sweeping and feathering angles referenced to this peak. Wing flex deformations, due to both gust impact and wing compliance, were greatest at the peak angular amplitudes of the strokes. These could have introduced some error into the estimation of wing angles, but were considered adequately accounted for due to the process of averaging over multiple strokes in the results. Supporting information on the setup and related experimental procedures can be found in \cite{jakobistrategies}.

\section{Results}
\subsection{Wing Motion In Unperturbed Flight}

\begin{figure}[hbt]
\centering
\includegraphics[width=1.1\columnwidth]{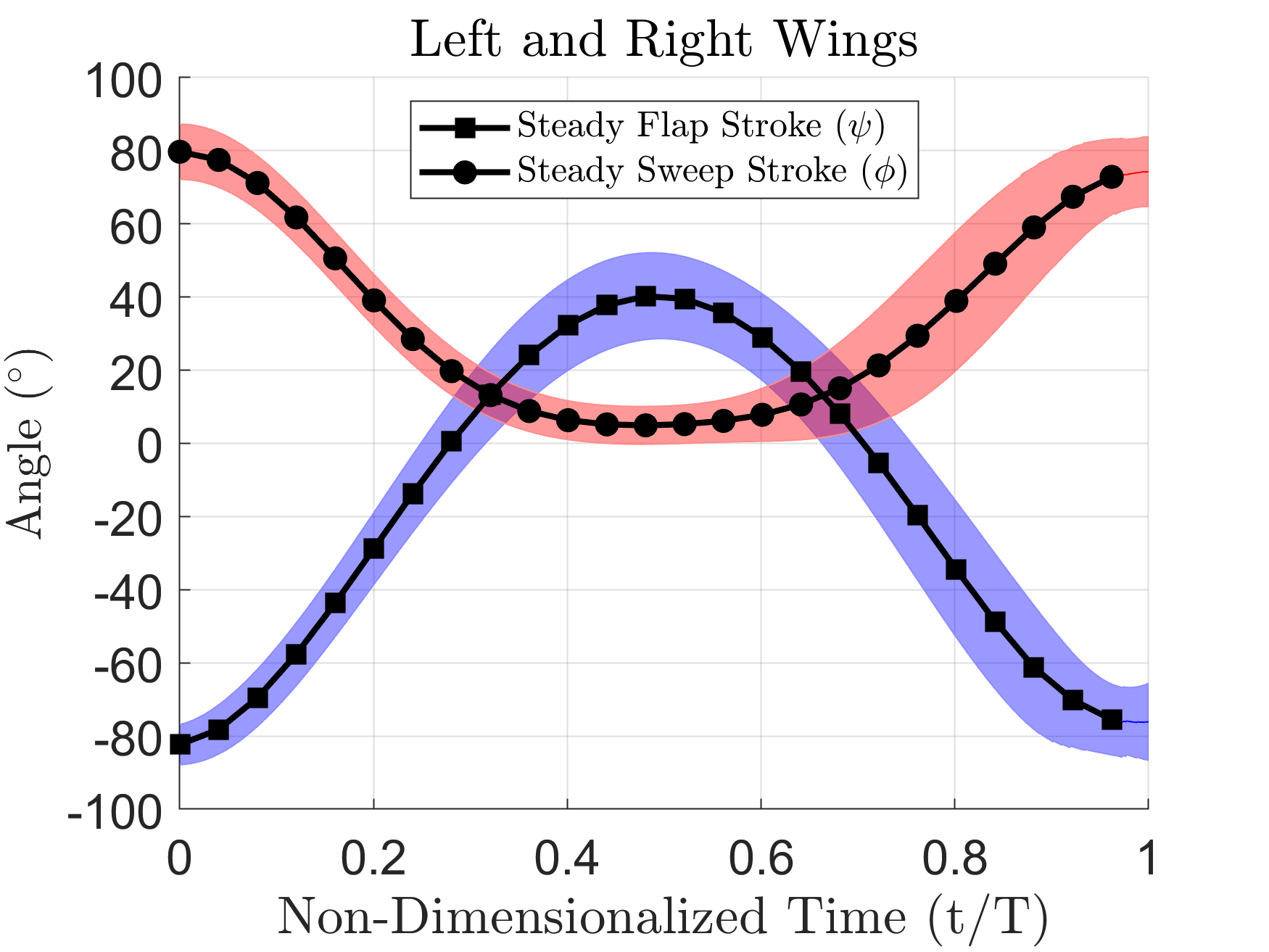}
\caption{Mean stroke sweep angles ($\phi$) and stroke flapping angles ($\psi$) (n = 108) plotted against non-dimensionalized stroke time (t/T). The start of the x-axis (t/T = 0) is the peak of the flap angle. Data from both the left and right wings were used to calculate the mean curves. The blue and red shaded regions indicate the standard deviation of the curves.}
\label{figure:steadykins}
\end{figure}

Wing flapping angles in steady flight were analyzed to establish a baseline for comparison with gust-induced changes. The black curve in Fig. \ref{figure:steadykins} represents the mean flap and sweep angles across a complete stroke, averaged from 108 wing stroke curves under steady conditions. The sweep angle had a stroke amplitude of $120.4^\circ \pm 12.2^\circ$, and the flap angle had an amplitude of $76.4^\circ \pm 9.50^\circ$ (Fig. \ref{figure:steadykins}), reflecting some variation due to mid-flight corrective behaviours. These findings are consistent with previous studies, such as in \cite{dudley1990mechanics}, which reported a maximum stroke amplitude of 115$^\circ$ to 125$^\circ$ in \textit{Bombus terrestris}.

\subsection{Wing Motion Adaptations in Different Gust Conditions}

\begin{figure}[!ht]
\centering
\includegraphics[width=1.1\columnwidth]{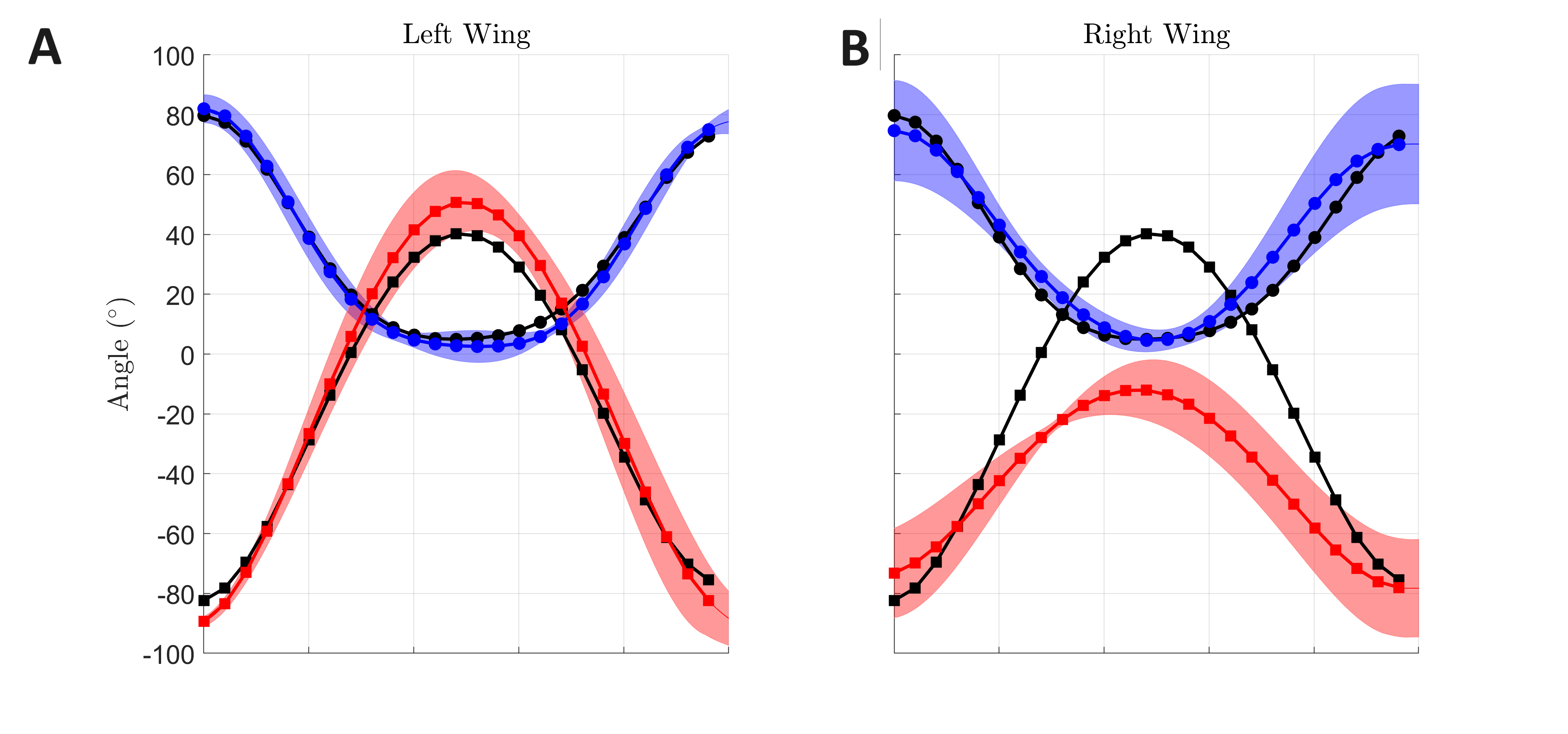}
\includegraphics[width=1.1\columnwidth]{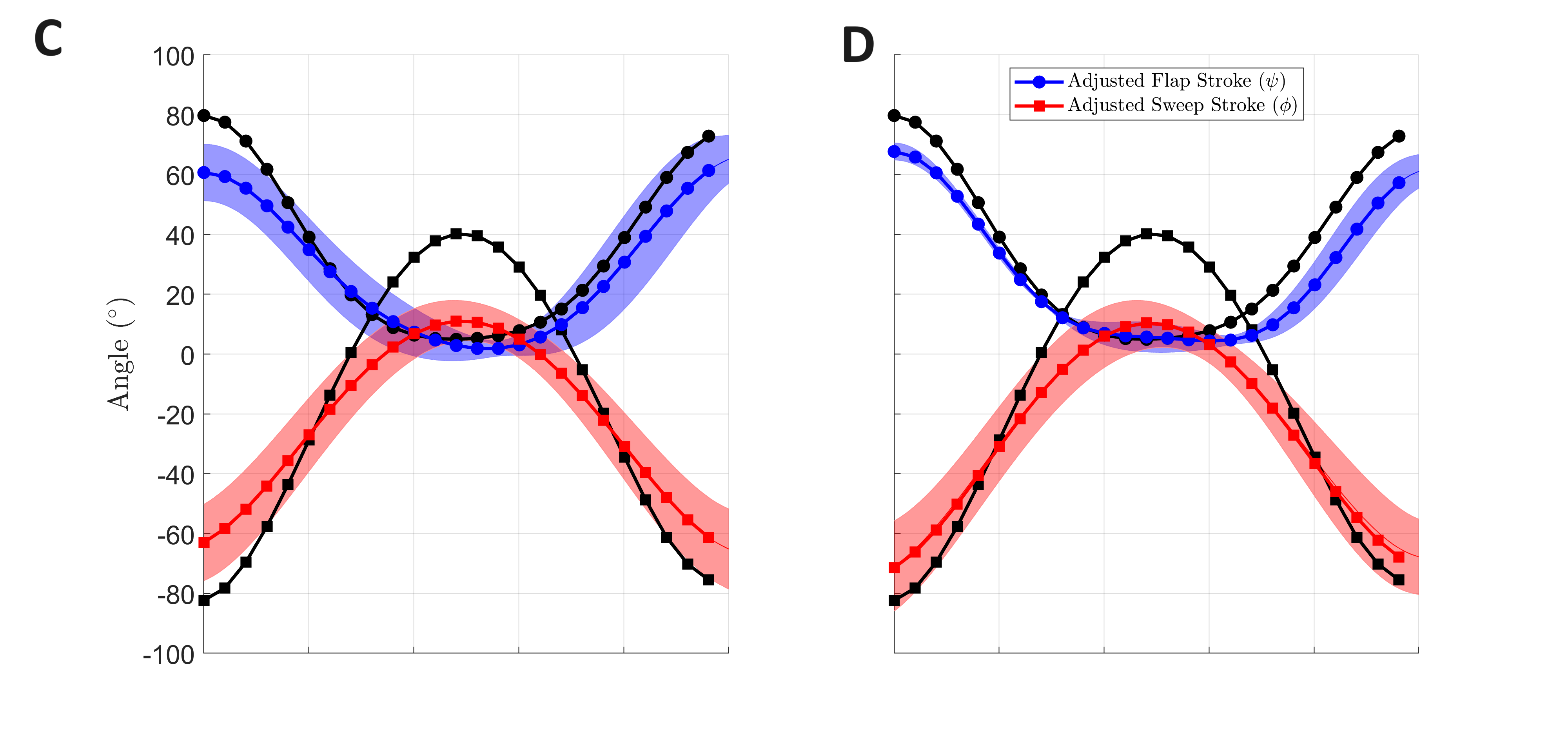}
\includegraphics[width=1.1\columnwidth]{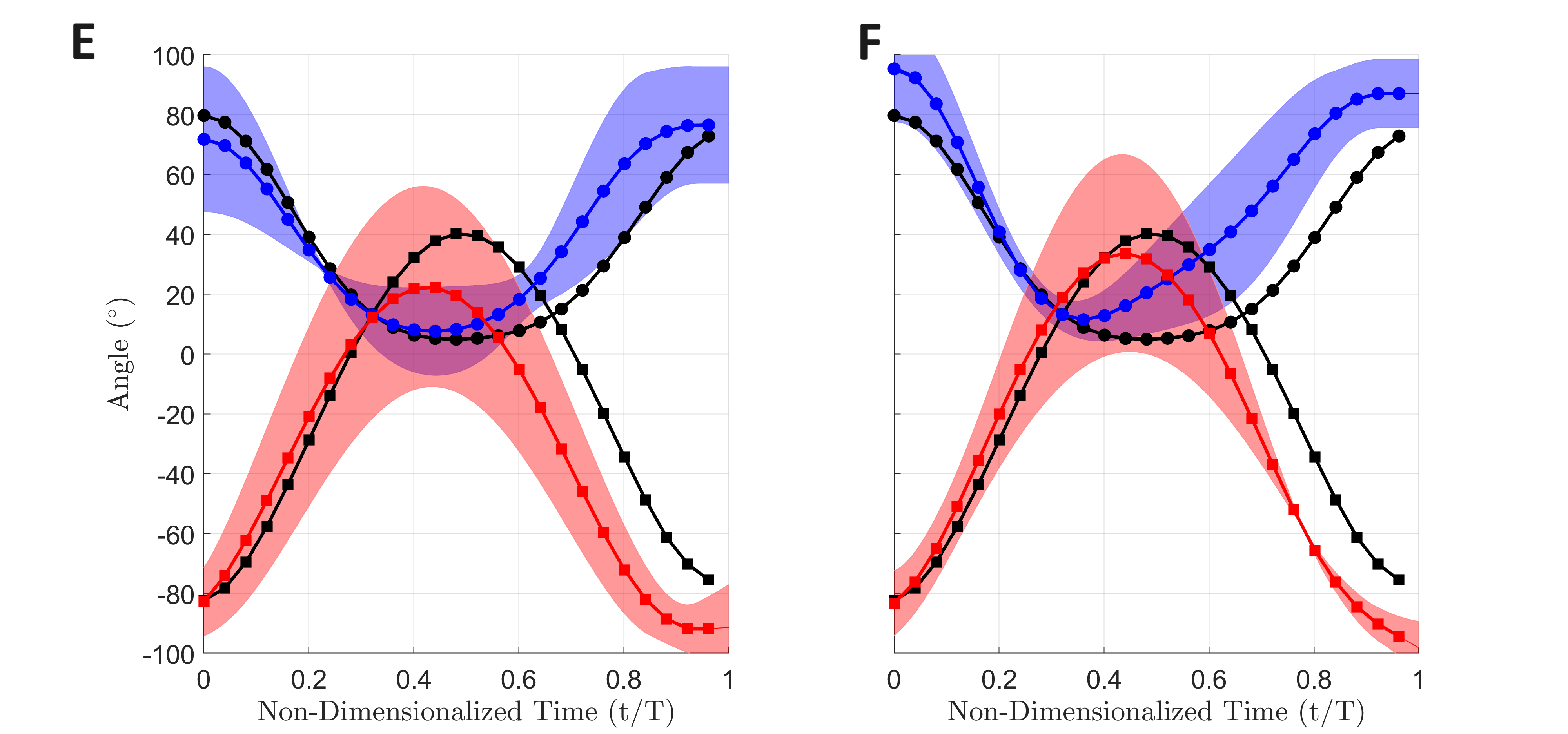}
\caption{Mean stroke sweep and flap angles in each of the three gust orientations plotted against non-dimensionalised stroke time (t/T). (A) and (B) show the left and right wings in the sideward gust, (C) and (D) present the data in the upward gust, and (E) and (F) show the data in the downward gust. The blue and red shaded
regions indicate the standard deviation of the curves. For reference, the mean flap and sweep curves for the steady case are shown in black.}
\label{figure:allkins}
\end{figure}

To capture the key changes in wing kinematics, we divided the flight into three distinct phases. The first phase, termed the steady phase, represents the period before the bees encountered the gust. The wing kinematics observed during this phase, as illustrated in Fig. \ref{figure:steadykins}, are included in Fig. \ref{figure:allkins} for direct comparison with the kinematic changes in the following phases. The second phase, termed the impulsive phase, occurred as the bees entered the gust but had not yet initiated any active responses. During this phase, all behaviours were passive or involuntary. The third phase, the recovery phase, began when the bees initiated counteractive manoeuvres in response to the gust. The wing kinematics during this phase represent the average stroke patterns as the bees attempted to return to their original flight path.

The bee body trajectory responses to gusts were published in a related study \cite{jakobi2018bees}. In the previous study, it was found that in flights through any gust direction, an impulsive orientation perturbation was generated by the gust in the direction the gust was pointing. For example, a sideward gust causes an impulsive negative yaw and a negative roll motion. When the bee engages in recovery, manoeuvres are undertaken against these perturbations. These related findings are summarised in Table \ref{tab:perturbations}.

\begin{table}[h!]
\centering
\caption{Summary of body orientation perturbations in gusts. The referenced perturbation directions are listed in Table \ref{tab:coordinates}.}
\begin{tabular}{lll}
\hline
\textbf{Gust Condition} & \textbf{Perturbation} & \textbf{Recovery Manoeuvre} \\
\hline
Sideward & -Yaw, -Roll &+Yaw, +Roll\\
Upward & +Pitch & -Pitch\\
Downward & -Pitch & +Pitch\\
\hline
\end{tabular}
\label{tab:perturbations}
\end{table}

\subsubsection{Sideward Gusts}

In sideward gusts, as bees penetrated the gust region, the stroke sweep angular range decreased to $55.1^\circ \pm 15.5^\circ$ for the wing closer to the gust (right wing), with the mean base value ($-79.5^\circ$) aligning with the approximate minimum of the sweep range observed during steady flight ($-80.7^\circ$, Fig. \ref{figure:steadykins}A). This posterior shift in the range of motion caused the right wing to remain behind the centre of gravity (CG), reducing the velocity and forces acting on the wing while maintaining a flapping frequency similar to that in steady flight. On the opposite side (the left wing), the stroke sweep angle increased to $130^\circ \pm 10.3^\circ$, extending beyond the full range of motion typical for these bees, as observed in the steady flight analysis. Despite these changes, the flap angle remained consistent for both wings, with similar angular ranges, reference positions, and timing, indicating no significant deviation from the steady flight condition (Fig. \ref{figure:allkins}A).

\subsubsection{Upward Gusts}

In response to upward gusts, the stroke sweep angle of both wings shifted posteriorly to a reduced range, with the minimum point at $-64^\circ$, closer to the minimum sweep observed during steady flight ($-80.7^\circ$). Despite this shift, flapping frequency and phase timing remained consistent with steady flight, resulting in lower velocities over the wings (reciprocating over a shorter angular range). The forward sweep angle peak decreased to $8^\circ$, confining most of the stroke range behind the CG at the onset of recovery. This adjustment effectively reduced the total stroke range to $72^\circ \pm 11.6^\circ$, nearly half of that observed during steady flight ($120.4^\circ \pm 12.2^\circ$). While the wing sweep was confined to this reduced posterior range, the flapping angles remained relatively consistent throughout the stroke cycle, peaking at $62.8^\circ \pm 6.9^\circ$, approximately $15^\circ$ lower than in steady flight ($76.4^\circ \pm 9.5^\circ$). Shifting the wings posteriorly relative to the bee's body while maintaining the angular flapping range likely enhanced stability by keeping the wings behind the CG, optimizing the reception of gust forces and increasing the distance between the gust-induced drag and the CG, contributing to a more stable, bottom-heavy system.

\subsubsection{Downward Gusts}

In response to downward gusts, wing motions closely resembled those observed in steady flight, with wings sweeping across the full angular range and occasionally extending beyond it ($118^\circ \pm 32.1^\circ$). Both wings exhibited a 20\% increase in stroke frequency, resulting in higher flow velocities over the wings as they reciprocated over a larger angular range. This increase in frequency was accompanied by greater variability in the strokes, with sporadic shifts in the anterior peak of one wing, leading to variations in the mean sweep and flap peaks ($\overline{\sigma_{\psi}} = 14.5^\circ$, $\overline{\sigma_{\phi}} = 20.0^\circ$ across both wings). These inconsistencies were likely associated with increased body roll and yaw oscillations during the perturbation. Additionally, the flapping angles were, on average, 12.7$^\circ$ higher than in the steady case across the stroke. The substantial increase in frequency and the tendency to sweep over a greater angular range likely represented an adaptive response by the bees to enhance flow velocity over the wings in reaction to the downward gust.

\section{Inferred Dynamics from Kinematic Adjustments}

In the experiments, bumblebees showed agility in their wing movements when responding to gust disturbances of different directions, dynamically articulating the wing towards unique configurations in each of the disturbance cases. Depending on the gust's direction, bees adjusted the mean position, range and frequency of their wing positioning relative to the body (through sweep and flap angles). Across all gust disturbances, sweep angle, or the anteroposterior configuration of the wing relative to the body was the main driver determining how bee attitudes changed in the gust. Variations in flapping angles were minimal compared to changes in sweep angles. While sweep and flap angles maintained coordinated adjustments throughout all responses to the gusts, changes in sweep angles did not result in corresponding changes in flapping angles within the same stroke, mostly maintaining its steady-flight articulation range across strokes. 

\begin{figure}[H]
\centering
\includegraphics[width=0.99\columnwidth]{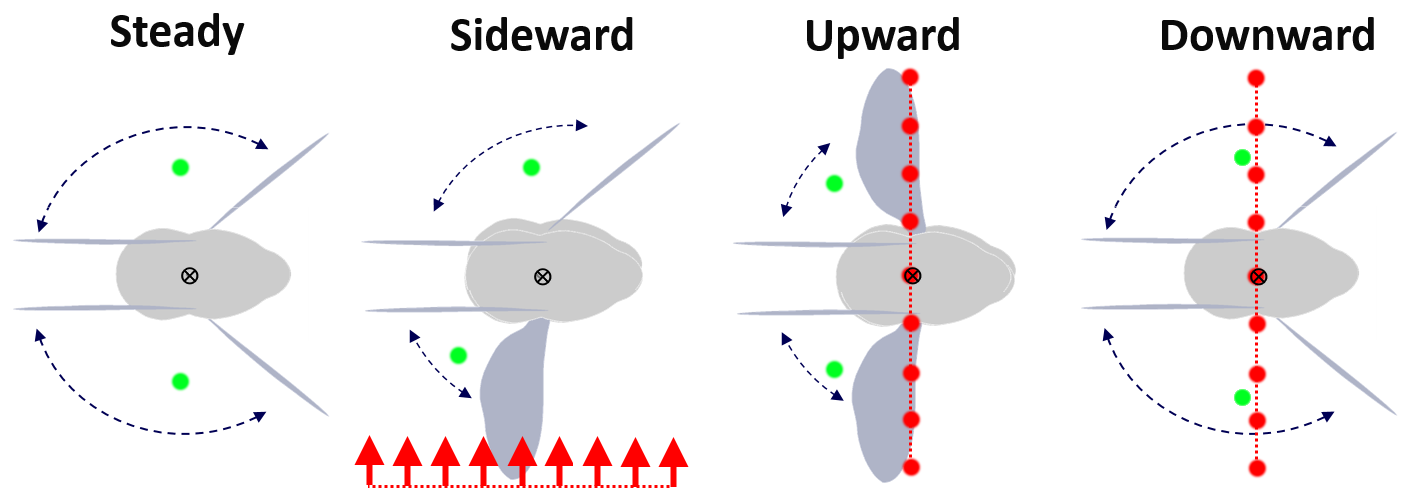}
\caption{Wing kinematics in response to different gust directions. The diagrams categorise the wing kinematic configurations observed in each gust test case. Red arrows indicate the direction of the gust, while curved blue arrows represent the angular range of wing motion. The AC is marked by a green dot, and the CG is denoted by a central dark dot. Note that initial impulsive perturbations caused by the gusts are not depicted; the bee’s position would actually be slightly displaced in the direction of the gust (red arrows) at the onset of these perturbations.}
\label{figure:compkins}
\end{figure}

The main wing adaptive configurations in each of the three different gust directions are displayed in Fig. \ref{figure:compkins}. In each case, the configuration generally appeared within 2-4 wing beats after experiencing the gust, and only lasted 1-2 wing beats before returning to a more steady stroke configuration (more similar to steady). This suggests that these brief, transient changes in the wing kinematics are critical in counteracting and recovering from gust-induced perturbations.

By considering the CG of bees to lie at the geometric centre of bees looking down (Fig. \ref{figure:compkins}), and assuming aerodynamic forces act at an aerodynamic centre (AC) located near the wing's centre of area, the basic dynamics of observed wing configuration changes (Fig. \ref{figure:allkins}) can be inferred. For sideward gusts, the wing closer to the gust shifted into a reduced sweep range. This effectively put the wing in a more posterior and dorsally raised position relative to the body for a greater proportion of the stroke. Using the illustration in Fig. \ref{figure:compkins} for reference, it can be seen that such a change would shift net wing forces further behind the CG (compared to the steady wing kinematics case), generating a restorative yaw moment toward the gust. Additionally, the reduced velocity on the wing closer to the gust results in an imbalance of upward force between the two wings, leading to a counteractive roll manoeuvre towards the gust. These roll and yaw corrections directly counteract the initial perturbation caused by the gust (Table \ref{tab:perturbations}) and facilitate a passive (utilising on forces from the gust) mechanism for handling sideward disturbances. Previous studies, such as \cite{viswanath2010effect}, suggest that bees tend to prefer frontal gusts; hence, it is plausible that bees initially correct perturbations towards the gust and, if the gust persists (as is often the case in natural environments), overcorrect to face into the gust, a behaviour observed in some of the slower bees in the experiments.

In response to upward gusts, both wings shifted into a sweep range configuration confined to the posterior region. Similar to the sideward gust scenario, but involving both wings, this adjustment positioned the wings more posteriorly and dorsally relative to the body for a greater portion of the stroke. Although the reduced wing velocity would typically result in lower force generation, the posterior positioning of both wings behind the CG is conducive to generating a counteractive pitch-down torque. The upward force from the gust flow enhances the wing-generated forces in the same direction, behind the CG, thereby amplifying the pitch-down moment and countering the nose-up perturbation induced by the gust (Table \ref{tab:perturbations}). By maintaining both wings in the posterior position behind the CG, the bees appear to optimize stability, effectively utilizing their wing position to harness the gust flow and augment the counteractive forces.

In the case of downward gusts, which induced an initial impulsive nose-down perturbation, it might be expected that the bees would shift their wing sweep range anteriorly to generate a pitch-up response, similar to the strategy employed in response to upward gusts. However, the bees instead increased their flapping frequency while maintaining a broad sweep range, likely to enhance lift and counteract the downward force. This adaptation appears to be an effort to sustain lift production and preserve flight stability under challenging conditions, particularly under the adverse conditions created by downward gusts, which are known to disrupt flows near flapping wings \cite{viswanath2010effect,jakobistrategies}. The findings suggest that downward gusts pose a particularly difficult scenario for flapping wing kinematics, as maintaining sufficient lift while counteracting the downward force necessitates more substantial adjustments in wing kinematics. This highlights a potential trade-off between generating the necessary upward forces and maintaining manoeuvrability during the disturbance, indicating that flapping wings may face inherent limitations in simultaneously achieving both objectives.

In two of the three gust orientations, one or both wings shifted their sweep range posteriorly, moving toward the lower half of the sweep spectrum and significantly reducing the median sweep angle. This shift likely enhances the stability of the bee by repositioning the aerodynamic forces relative to the CG, making the bee's response to disturbances more effective. this simple change for each wing can immediately provide stability enhancing effects, making it better for flight.

\section{Conclusion}

This study explored the adaptive wing kinematics of bumblebees under gust disturbances. The primary adaptive strategy observed was the posterior shift in wing sweep angles, which varied with gust direction. In sideward gusts, this shift in the wing closest to the gust enhanced lateral stability, while upward gusts induced a coordinated posterior shift in both wings to counteract pitch-up manoeuvres. For downward gusts, bees increased flapping frequency to maintain lift, though this may have limited their ability to counteract pitch-down perturbations.

These findings demonstrate the sophisticated control mechanisms employed by bumblebees to maintain stability in turbulent conditions. The posterior wing shift, a defining kinematic adjustment seen in the results, could enhance flapping manoeuvrability by dynamically adjusting wings with a similar shift during disturbances. Incorporating mechanisms to modulate wing position or balance force generation between wings could greatly improve MAV control and resilience in turbulent environments.

While this paper offers a qualitative analysis, further detailed statistical analyses will be presented in a forthcoming full paper, which will delve into the statistical significance of these findings and their implications for MAV design.


\bibliographystyle{unsrt}
\bibliography{imav_bibliography}


\end{document}